\documentclass{aa}

\usepackage[varg]{txfonts}

\usepackage{natbib}
\bibpunct{(}{)}{;}{a}{}{,} 
\usepackage{hyperref}
\usepackage{graphicx}
\usepackage{url}
\usepackage{amsmath}
\usepackage{amssymb} 
\usepackage{amsfonts}
\usepackage{aas_macros}
\usepackage{fancyvrb}
\usepackage{listing}
\usepackage{color}
\usepackage{pygments_style}

\hypersetup{colorlinks=true,citecolor=blue}

\newcommand{\FoMToT}{\mathrm{FoM}_{\mathrm{TOT}}}
\newcommand{\FoMDE}{\mathrm{FoM}_{\mathrm{DETF}}}

\title{3D galaxy clustering with future wide-field surveys: Advantages of a spherical Fourier-Bessel analysis} 
\titlerunning{Spherical 3D analysis of galaxy clustering}

\author{F. Lanusse \inst{1}\thanks{francois.lanusse@cea.fr}, A. Rassat \inst{2} \thanks{anais.rassat@epfl.ch},  J.-L. Starck \inst{1}}

\institute{$^1$ Laboratoire AIM, UMR CEA-CNRS-Paris, Irfu, SAp, CEA Saclay, F-91191 GIF-SUR-YVETTE CEDEX, France.\\$^2$ Laboratoire d'Astrophysique, Ecole Polytechnique F\'ed\'erale de Lausanne (EPFL), Observatoire de Sauverny, CH-1290, Versoix, Switzerland.}

\begin{document}

\abstract
{
Upcoming spectroscopic galaxy surveys are extremely promising to help in addressing the major challenges of cosmology, in particular in understanding the nature of the dark universe. The strength of these surveys, naturally described in spherical geometry, comes from their unprecedented depth and width, but an optimal extraction of their three-dimensional information is of utmost importance to best constrain the properties of the dark universe.
}
{
Although there is theoretical motivation and novel tools to explore these surveys using the 3D spherical Fourier-Bessel (SFB) power spectrum of galaxy number counts $C_\ell(k,k^\prime)$, most survey optimisations and forecasts are based on the tomographic spherical harmonics power spectrum $C^{(ij)}_\ell$. The goal of this paper is to perform a new investigation of the information that can be extracted from these two analyses in the context of planned stage IV wide-field galaxy surveys.
}
{
We compared tomographic and 3D SFB techniques by comparing the forecast cosmological parameter constraints obtained from a Fisher analysis. The comparison was made possible by careful and coherent treatment of non-linear scales in the two analyses, which makes this study the first to compare 3D SFB and tomographic constraints on an equal footing. Nuisance parameters related to a scale- and redshift-dependent galaxy bias were also included in the computation of the 3D SFB and tomographic power spectra for the first time.
}
{
Tomographic and 3D SFB methods can recover similar constraints in the absence of systematics.  This requires choosing an optimal number of redshift bins for the tomographic analysis, which we computed to be $N=26$ for $z_{med} \simeq 0.4$, $N=30$ for $z_{med}\simeq1.0,$ and $N=42$ for $z_{med} \simeq 1.7$.
When marginalising over nuisance parameters related to the galaxy bias, the forecast 3D SFB constraints are less affected by this source of systematics than the tomographic constraints.  In addition, the rate of increase of the figure of merit as a function of median redshift is higher for the 3D SFB method than for the 2D tomographic method. 
}
{
Constraints from the 3D SFB analysis are less sensitive to unavoidable systematics stemming from a redshift- and scale-dependent galaxy bias. Even for surveys that are optimised with tomography in mind, a 3D SFB analysis is more powerful.  In addition, for survey optimisation, the figure of merit for the 3D SFB method increases more rapidly with redshift, especially at higher redshifts, suggesting that the 3D SFB method should be preferred for designing and analysing future wide-field spectroscopic surveys. CosmicPy, the Python package developed for this paper, is freely available at \url{https://cosmicpy.github.io}.
}

\maketitle

\section{Introduction}
Understanding the nature of the dark universe is one of the fundamental challenges of modern cosmology today.  Galaxy clustering - the statistical analysis of the spatial distribution of galaxy number counts - has been identified as one of the most promising probes available to explore this \citep{Peebles:1980,DETF,WGFC}, with spectroscopic surveys being particularly useful in probing both tangential and radial modes in the Universe.

Galaxy number counts have been extensively studied with current and planned future surveys, and the analysis can be performed in various spaces, for example, Fourier space \citep{Seo:2003Eis,Seo:2007Eis}, configuration space \citep{Eisenstein:2005,baoconfig1,baoconfig2}, and spherical harmonic space \citep[e.g.,][]{Dolney:2006,DK:2012}. For future wide-field spectroscopic surveys, the galaxy field will cover large areas on the sky so that an analysis in spherical space provides a natural decomposition for certain physical effects as well as selection effects. For wide-field spectroscopic surveys, the depth of the survey means that a 3D spherical Fourier-Bessel (SFB) analysis is the most natural to perform \citep{Fisher:1995,Heavens:1995,Rassat:2012bao}.  

Previous SFB analyses of the local Universe \citep[e.g.,][]{Erdogdu:2006dv,Erdogdu:2005wi} used relatively small data sets, where straightforward summation methods were sufficient to measure the SFB coefficients. Today, novel numerical methods for 3D spherical analysis are available \citep{3DEX,Lanusse:2011} to prepare for future wide-field surveys that will map the large-scale structure of the Universe with a large number of galaxies. The 3D SFB analysis can also be applied to other probes, for instance weak-lensing \citep{Heavens:2003,CHK:2005,Kitching:2008,KHM:2010,Schafer:2014,3dwl:bao,3dwl:2014} and the integrated Sachs–Wolfe effect \citep[e.g.][]{ISW3D}, which will be crucial for high-precision probe combinations.

Even given the existing 3D SFB tools and the theoretical motivation for this approach, most existing forecasts and survey optimisation for future wide-field surveys focus on a tomographic analysis, that is, one where the survey is split into redshift bins, and 2D spherical harmonic auto- and cross-power spectra $C^{ij}(\ell)$ are measured  \citep[e.g.,][]{Redbook,DK:2012}. One of the advantages of a  tomographic spherical harmonics analysis is that there are several available codes to rapidly calculate the tomographic spectra, either for galaxy correlations or for other complementary probes \cite[e.g.,][]{class:gg,icosmo1}; another advantage is that it is straightforward to convert survey observables ($\theta, \phi, z$) into a power spectrum measurement without any assumption of cosmological parameters, while the 3D SFB analysis requires an assumption about the fiducial cosmology to translate the observables into the 3D SFB spectrum, which has a dependence of wavenumber $k$. However, in the tomographic analysis, some of the radial information may be lost as a result of redshift binning, while the 3D SFB analysis potentially uses the entire 3D information, especially for a spectroscopic survey. With this in mind, a natural hypothesis is that a 3D SFB spectroscopic analysis might extract more information than a tomographic one.

Several studies have already investigated this, for example, \cite{Clzz} and \cite{Asorey:2012} found that a tomographic analysis returned equivalent or better constraints than a 3D Fourier power spectrum analysis. They concluded that the tomographic approach should be preferred as it avoids the need to assume a particular cosmology to convert redshifts into comoving distances and simplifies the  combination with other probes such as weak-lensing. Nevertheless, they both acknowledged that for a spectroscopic survey the tomographic analysis would require a large number of redshift bins to recover the full 3D information, which is limited by shot noise problems. For the first time, \cite{Nicola:2014} compared the tomographic analysis to a 3D SFB analysis and found the tomographic constraints to be superior, but they still noted that the 3D SFB approach was stable with regard to the choice of fiducial cosmology for the necessary conversion from redshift to comoving distance. However, their treatment of the non-linear scale cut-off used in the Fisher matrix comparison is not equivalent between the tomographic and the 3D SFB analysis. While the 3D power spectra is cut off at a physical scale ($k$) corresponding to nonlinear effects, the tomographic power spectra are truncated at fixed arbitrary angular scales. This ignores the interplay between the redshift of the tomographic bins and the wavenumber of the SFB spectrum. As a result, the non-linear cut-off in \cite{Nicola:2014} does not allow a fair comparison between 2D and 3D methods, which means understanding the strength of each method is still an open question. We address this question by carefully excluding non-linear scales.

Understanding how best to extract information for a 3D galaxy survey is of utmost importance to address the fundamental questions in modern cosmology today, and also to ensure that future planned surveys are efficiently analysed as well as optimised. To address this pressing question, we propose here a new investigation of the information that can be extracted from a spectroscopic galaxy survey by tomographic vs. 3D SFB analysis. Our approach focuses on the seven common parameters that are currently used in wide-field survey optimisation and planning, that is, on $\vec{\theta}=\{\Omega_m, h, w_0, w_a, \sigma_8, \Omega_b, n_s\}$, while putting forward a coherent approach regarding the exclusion of non-linear scales for both the 2D and 3D methods for the first time. In addition, we investigate for the first time how tomographic and 3D SFB methods are affected by nuisance parameters related to the galaxy bias, which we allow to be both redshift- and scale-dependent. However, we do not include redshift space distortions (RSD) or relativistic effects in our study. Including RSDs, which will be present in the data, provides an additional probe, which improves constraints. Although a prescription for RSDs in SFB space exists \citep{Heavens:1995}, as a first approach, we do not include them here in either the tomographic or the SFB  analysis
to ensure that we compare like with like. Their impact should nonetheless be assessed, which we plan to do in a future work.

Finally, in the spirit of reproducible research, we make available all tomographic and 3D SFB codes used for this analysis, along with the scripts to reproduce our results.

Our paper is structured as follows: in Sect. \ref{sec:theory}, we briefly review the theory behind the statistical analysis of galaxy number counts, including the prescription for the tomographic analysis and the 3D SFB. In Sect. \ref{sec:Forecasting}, we provide
an overview of the Fisher matrix forecasting-approach that we used to compare the relative constraining power of each method, and include the description of the future spectroscopic wide-field survey for which we calculate forecasts, of the question of non-linear scale treatment, and galaxy bias nuisance parameters. In Sect. \ref{sec:results}, we present the comparison between the constraining power of the 3D SFB and tomographic methods and investigate how this comparison holds in the presence of galaxy bias nuisance parameters. We also determine how this affects a future wide-field survey optimisation. In Sect. \ref{sec:ccl}, we present our conclusions in the context of high-precision cosmology with future wide-field surveys. 

\section{Theory}
\label{sec:theory}
In this first section we describe the formalism behind the analysis of galaxy clustering in the context of a spectroscopic survey. We present the two methodologies compared in this work, one based on a tomographic analysis of angular correlations, the other based on the correlations of the 3D expansion of the galaxy field on a spherical Fourier-Bessel basis.

\subsection{Galaxy and matter fields}

In a galaxy survey, the quantity observed is the galaxy number density $n(\mathbf{r} = (r,\theta,\varphi)),$ which can be defined in terms of the galaxy overdensity $\delta_g$ through
\begin{equation}
        n(\mathbf{r}) = \bar{n}(r) ( 1 + \delta_g(\mathbf{r}, z(r))) \;,
        \label{eq:density}
\end{equation} 
where $\bar{n}(r)$ is the mean number density of observed galaxies at comoving distance $r$. In this expression, the time dependence of the observed overdensity as a function of comoving distance is made explicit through the $z(r)$ relation. The mean number density $\bar{n}(r)$ can be expressed in terms of the survey selection function $\phi(r)$ as
\begin{equation}
        \bar{n}(r) = \phi(r) \bar{n} = \frac{N}{V} \phi(r) \;,
\end{equation} 
with $\bar{n}$ the mean number density of observed galaxies, $N$ the total number of observed galaxies, and $V$ the volume of the survey that fulfils $V = \int \phi(r) \mathrm{d}\mathbf{r}$. Note that in the general case, the selection function has both an angular and a radial dependence (see Sect.~\ref{sec:fsky}), but in this work, we did not consider the impact of an angular mask and only account for partial coverage of the sky through a multiplicative $f_{sky}$ factor.

In expression \eqref{eq:density}, the time (or redshift) dependence of the galaxy overdensity is due to the growth of structure and the evolution of galaxy bias with respect to the matter density field with time. Following the approach of \citet{Rassat:2012bao}, in the linear regime this dependence on redshift can be separated in the form of growth and bias prefactors,
\begin{equation}
        \delta_g(\mathbf{r},z(r)) = b(r, k) D(r) \delta(\mathbf{r}) + \epsilon(\mathbf{r}) \;, \label{eq:galaxy_overdensity_to_matter_overdensity}
\end{equation}
where $b(r, k)$ is a bias with a possible scale dependence, $D(r)$ is the growth factor, $\delta(\mathbf{r}) = \delta(\mathbf{r}, z = 0)$ is the matter overdensity field at present day, and $\epsilon(\mathbf{r})$ is a Poisson noise term arising from the discrete nature of the observed galaxy number density. As in \citet{Rassat:2008bao}, we considered the linear relation \eqref{eq:galaxy_overdensity_to_matter_overdensity} to hold in the standard cosmological model on large scales up to a redshift-dependent $k_{\max}(z)$ with $k_{\max}(z = 0) \simeq 0.12 h$Mpc$^{-1}$ and $k_{\max}(z = 2) \simeq 0.25 h$Mpc$^{-1}$. We then proceeded to define a modified selection function that includes the effects of bias and growth in the linear regime,
\begin{equation}
        \phi^{\mathrm{evol}} = b(r, k) D(r) \phi(r) \;.
        \label{eq:modified_selection}
\end{equation}
Using this modified selection function, the observed galaxy density can now be expressed directly as a function of the true matter overdensity at present time:
\begin{equation}
        \frac{n(\mathbf{r})}{\bar{n}} = \phi(r) + \phi^\mathrm{evol}(r, k)\delta(\mathbf{r}) + \phi(r) \epsilon(\mathbf{r}) \;.
        \label{eq:density_to_overdensity}
\end{equation}

\subsection{Tomographic analysis of galaxy clustering}

In the tomographic analysis, the survey is decomposed into spectroscopic redshift bins from which are computed classical angular correlation functions. The angular number density for one spectroscopic bin $(i)$ limited between $z_{\min}^{(i)}$ and  $z_{\max}^{(i)}$ is defined as
\begin{equation}
        n^{(i)}(\mathbf{\theta}) = \bar{n}^{(i)} \left(1 + \delta^{(i)}(\mathbf{\theta})\right) = \int_{z_{\min}^{(i)}}^{z_{\max}^{(i)}} n(z, \mathbf{\theta}) \mathrm{d} z \;,
\end{equation}
where $\bar{n}^{(i)}$ is the average galaxy number density per steradians in tomographic bin  $(i)$ and $\delta^{(i)}(\mathbf{\theta})$ is the angular galaxy overdensity in bin $(i)$. Expanding the angular overdensity in spherical harmonics yields
\begin{equation}
        n_{\ell m}^{(i)} = \int n^{(i)}(\mathbf{\theta}) Y_{\ell m}^*(\mathbf{\theta}) \mathrm{d} \mathbf{\theta} \;,
\end{equation}
From this spherical harmonics expansion, the tomographic angular correlation functions between bins $i$ and $j$, noted $\overline{C}^{(i j)}_\ell$, is defined for $\ell \geq 1$ as
\begin{align}
        {\overline{C}^{m m^\prime}_{\ell \ell^\prime}}^{(i j)} &\equiv \frac{1}{\bar{n}^{(i) 2}} <n_{\ell m}^{(i)} n_{\ell^\prime m^\prime}^{(j) *} > \;, \\
        &= \left( C^{(i j)}_\ell + \frac{\delta^K_{i j}}{ \bar{n}^{(i)}} \right) \delta_{\ell \ell^\prime}^K \delta^K_{m m^\prime} \;,
        \label{eq:tomopowspec}
\end{align}
where $\delta^K$ is the Kronecker symbol. In the last equation, the first term $C^{(i j)}_\ell$ is the contribution from galaxy clustering and the second term $\frac{1}{\bar{n}^{(i)}}$ is the contribution from shot noise, which only affects the auto-correlation power spectra. Note that different angular modes are predicted to be uncorrelated in linear theory for a Gaussian random field; these can become correlated as a result of non-linearities or lack of full-sky coverage, effects that we did not consider in this work. Formally, the correlation functions $C_\ell^{(ij)}$ are related to the matter power spectrum $P(k)$ at $z = 0$, in the linear regime, according to
\begin{align}
C^{(i j)}_{\ell} &= \frac{2}{\pi} \int  \mathrm{d}k P(k) k^2 \int w^{(i)}_\mathrm{evol}(r,k) j_\ell(k r) \mathrm{d}r \nonumber \\
 & \quad \times \int w^{(j)}_\mathrm{evol}(r^\prime,k) j_\ell(k r^\prime) \mathrm{d}r^\prime \;,
\end{align}
where $w_{\mathrm{evol}}^{(i)}$ is a window function for bin $(i),$ which includes the effects of spectroscopic selection, linear growth, and bias:
\begin{equation}
        w^{(i)}_{\mathrm{evol}} = \phi^{\mathrm{evol}}(r,k) s^{(i)}(r) \;,
\end{equation}
with $\phi^{\mathrm{evol}}$ is the modifier selection function including growth and bias introduced in Eq. \eqref{eq:modified_selection} and $s^{(i)}$ is the spectroscopic selection function that defines the redshift bin $i$ , that is,  $s^{(i)}(z) = 1$ if $z \in [z^{(i)}_{\min}, z^{(i)}_{\max}]$, $s^{(i)}(z) = 0$ otherwise.

This expression is the full general expression of the tomographic angular power spectrum. However, it is common to evaluate the angular power spectrum through the well-known Limber approximation. To the first order \citep{Loverde:2008limber}, the Limber approximation applied to the previous equation yields\begin{multline}
        C^{(i j)}_{\ell |_{\mathrm{ Limber}}} = \int \frac{\mathrm{d}r}{r^2} P\left(\frac{\ell + 1/2}{r}\right) w^{(j)}_\mathrm{evol}\left(r,\frac{\ell + 1/2}{r}\right) \\
        \times \quad w^{(i)}_\mathrm{evol}\left(r,\frac{\ell + 1/2}{r}\right) \;.
\end{multline}
The Limber approximation holds to very good accuracy for the auto-correlations under the assumption that the bin window functions do not vary too rapidly or that the overlap between bins is not too small.

\subsection{3D spherical Fourier-Bessel analysis of galaxy clustering}
\label{sec:3DSFB}

The spherical Fourier-Bessel transform of the galaxy number density $n(\mathbf{r})$ is defined as\begin{equation}
        n_{l m}(k) = \sqrt{\frac{2}{\pi}} \int n(\mathbf{r}) k j_\ell(k r) Y_{l m}^*(\theta,\varphi) \mathrm{d}\mathbf{r} \;,
        \label{eq:SFB}
\end{equation}
where $j_\ell$ are spherical Bessel functions, $Y_{\ell m}$ are spherical harmonics, $\ell$ and $m$ are multipole moments, and $k$ is the wavenumber. Note that in this work we follow the orthonormal convention for the SFB, as in \citet{Rassat:2012bao}, \citet{Fisher:1995}, or \citet{Pratten:2013}. From the SFB coefficients $n_{\ell m}(k)$, the number density can be recovered through the inverse SFB transform as
\begin{equation}
        n(r, \theta, \varphi) = \sqrt{\frac{2}{\pi}} \sum\limits_{\ell, m} \int n_{\ell m}(k) k j_\ell(k r) \mathrm{d}k Y_{\ell m}(\theta,\varphi) \;.
        \label{eq:Inv_SFB}
\end{equation}
Although the SFB expansion is performed in comoving space, in practice, the galaxy number density is only observed in redshift space. This means that a fiducial cosmology has
to be assumed to relate observed redshift and comoving distance of the galaxies in the survey. To distinguish between true comoving distance $r$ and estimated comoving distance, we introduce the notation
\begin{equation}
        \tilde{r} \equiv r_{|_\mathrm{fid}}(z)  \;.
\end{equation}
When the fiducial cosmology exactly corresponds to the true cosmology, $\tilde{r} = r,$ but in general, this is not the case. The importance of making this distinction has been stressed in \citet{Heavens:2006}, especially when constraining dark energy parameters, which are very sensitive to the $r(z)$ relation.

For multipoles of order $\ell \ge 1$, the 3D SFB spectrum of the observed galaxy density can be expressed in the form
\begin{align}
        \overline{C}_{\ell \ell^{\prime}}^{m m^\prime}(k, k^\prime) &\equiv \frac{1}{\bar{n}^2} <n_{\ell m} (k) n_{\ell^\prime m^\prime}^* (k^\prime) > \;, \\
          &= \left( C_\ell(k, k^\prime) + N_\ell(k,k^\prime) \right) \delta_{\ell \ell^\prime}\delta_{m m^\prime} \;. \label{eq:SFBpowspec}
\end{align}
This expression can be directly compared to the definition of the tomographic power spectra in Eq. \eqref{eq:tomopowspec}. Just like in the tomographic case, different angular multipoles are not correlated when an angular mask is neglected. In this expression, the signal power spectrum $C_{\ell}(k, k^\prime)$ takes the form (see \cite{Rassat:2012bao} for this exact prescription or \cite{Heavens:1995})
\begin{equation}
        C_\ell(k, k^\prime)= \left(\frac{2}{\pi}\right)^2 \int k^{\prime \prime 2} P(k^{\prime \prime}) W_\ell^{\mathrm{evol}}(k,k^{\prime \prime}) W_\ell^{\mathrm{evol}}(k^\prime,k^{\prime \prime}) \mathrm{d}k^{\prime \prime} \;,
\end{equation}
where the following window function includes the effects of linear growth and bias and the fiducial redshift-comoving distance relation:
\begin{equation}
        W_\ell^{\mathrm{evol}}(k,k^{\prime \prime}) = k \int \phi^\mathrm{evol}(r, k^{\prime \prime}) j_\ell(k \tilde{r}) j_\ell(k^{\prime \prime} r)  r^2 \mathrm{d}r \;.
\end{equation}
The noise covariance matrix can be expressed as
\begin{equation}
        N_\ell(k,k^\prime)  = \frac{2 k k^\prime}{\bar{n}\pi} \int \phi(r) j_\ell(k \tilde{r}) j_\ell(k^\prime \tilde{r})  r^2 \mathrm{d}r \;.
\end{equation}
This expression is equivalent to that used in \citet{Desjacques:bao}, and a derivation can be found in Appendix~\ref{sec:ShotNoise}.

When considering a realistic galaxy survey with finite depth, the observed galaxy number density vanishes above a given $r_{\max}$ and fulfils
\begin{equation}
        \forall (\theta, \phi), \quad n(r_{\max}, \theta, \phi) = 0 \;.
\end{equation} 
Under this boundary condition, the spherical Fourier-Bessel transform can be inverted from discretely sampled coefficients $n_{\ell m}(k_{\ell n})$ and \eqref{eq:Inv_SFB} becomes
\begin{equation}
        n(r, \theta, \phi) = \sum\limits_{\ell, m, n} \kappa_{\ell n} n_{\ell m}(k_{\ell n}) k_{\ell n} j_\ell(k_{\ell n} r) Y_{\ell m}(\theta,\phi) \;,
\end{equation}
where the discrete wavenumbers $k_{\ell n}$ are defined in terms of the zeros of the spherical Bessel function $q_{\ell n}$ as 
\begin{equation}
        k_{\ell n} = \frac{q_{\ell n}}{r_{\max}} \;,
\label{eq:scale_sampling}
\end{equation}
and the normalisation factors $\kappa_{\ell n}$ are defined as $\kappa_{\ell n} = \frac{\sqrt{2 \pi} r_{\max}^{-3}}{j_{l+1}^2(q_{l n})}$ \citep{Fisher:1995}.

In the context of Fisher matrix forecasting, the main consequence of this discretisation is that it imposes a discrete sampling of the SFB spectrum that can be represented in matrix form $C_\ell(n, n^\prime) = C_\ell(k_{\ell n},  k_{\ell n^\prime})$ without loss of information.

\subsection{Effect of partial sky coverage}
\label{sec:fsky}

So far, we have assumed complete sky coverage. However, obscuration and confusion due to our own galaxy means that only a portion of the sky is observable in practice. This effect can be modelled in a similar way for both tomographic and SFB derivations by applying an angular weighting function $M(\theta,\phi)$ to the galaxy density field, for instance with $M(\theta,\phi) = 0$ in masked areas and $M(\theta, \phi) = 1$ otherwise. The effect of such a mask on angular power spectra is well known and results in a coupling of angular modes that would otherwise remain uncorrelated. Formally, the signal part of both tomographic and SFB power spectra becomes
\begin{align}
        C_{\ell \ell^\prime}^{m m^\prime (i j)} &= \sum\limits_{\ell'' m''} M_{\ell m \ell'' m''} M_{\ell' m' \ell'' m''} C_{\ell''}^{(i j)} \;, \\
                C_{\ell \ell'}^{m m'} (k, k') &= \sum_{\ell'' m''} M_{\ell m \ell'' m''} M_{\ell' m' \ell'' m''} \iint  K_{\ell \ell''}(k,k_1) K_{\ell' \ell''}(k',k_2) \nonumber  \\
                &\quad \times \quad  C_{\ell''}(k_1, k_2) \quad \mathrm{d}k_1 \mathrm{d}k_2 \;,
\end{align}
where $M_{\ell m \ell'' m''}$ is an angular coupling kernel defined in terms of the angular mask as
\begin{equation}
        M_{\ell m \ell' m'} = \int_\Omega Y_{\ell^\prime m^\prime}(\Omega) M(\Omega) Y_{\ell m}^*(\Omega) \;,
\end{equation}
and $K_{\ell \ell'}(k, k')$ is a wavenumber coupling kernel defined as
\begin{equation}
        K_{\ell \ell'}(k, k') =  \frac{2}{\pi} k^{\prime 2} \int_r j_{\ell}(k r) j_{\ell'}(k' r) r^2 dr \;.
\end{equation}
The noise part of the power spectra that is the result
of shot noise can be approximated as only affected by a simple area scaling \citep{3dwl:2014}. Note that the coupling matrix $K_{\ell \ell'}(k, k')$ reduces to a Dirac delta function when $\ell = \ell'$ thanks to the orthogonality of the spherical Bessel functions. In both cases, the mask will induce a coupling of angular modes. In the SFB case, this also means that the coupling kernels $K_{\ell \ell'}(k, k')$ can no longer be considered as Dirac delta functions and induce an additional coupling between different wavenumbers.

In practice, the impact of the mask can conveniently be taken into account using the pseudo-$C_\ell$ methodology, which is well known for studies of the cosmic microwave background \citep{Hivon:2002}. In the tomographic as well as the SFB analysis, the pseudo-$C_\ell$ estimator can be linked to the theoretical $C_\ell$ power spectrum using either a 2D or 3D mixing matrix:
\begin{align}
        <\widetilde{C}_\ell(k_{\ell n}, k_{\ell n'})> &= \sum\limits_{\ell' n_1 n_2} M^{3D}_{\ell \ell' n n_1 n'  n_2} C_{\ell'}(k_{\ell' n_1}, k_{\ell' n_2}) \;, \\
        <\widetilde{C}_\ell^{(i j)}> &= \sum\limits_{\ell \ell'} M^{2D}_{\ell \ell'} C_{\ell'}^{(i j)} \;.
\end{align} 
A derivation of the 2D mixing matrix can be found in \cite{Hivon:2002}, while the 3D matrix for the galaxy clustering SFB power spectrum is derived in \cite{Pratten:2013}.

For the purpose of this paper, it is important to point out that the effect of the mask can be taken into account in a similar way using pseudo-$C_\ell$s for the two methodologies explored here. Consequently, in a likelihood analyses using these expressions for the measured power spectra, the effect of the mask should be equivalent for the tomographic and SFB approaches. Therefore we only took partial sky coverage  through the common $f_{sky}$ scaling factor into account here for simplicity. This is standard practice for Fisher matrix analyses.

\section{Forecasting cosmological constraints}
\label{sec:Forecasting}

\subsection{Fisher matrix forecasting}

Expected cosmological constraints using the two different analysis techniques introduced in the previous section can be estimated with the Fisher matrix formalism \citep*{Tegmark:1997th}. The Fisher information matrix provides a lower bound on the expected errors on cosmological parameters under the assumption that the likelihood can be approximated by a Gaussian at its peaks. It is formally defined as the expectation value of the second derivative of the logarithmic likelihood with respect to the parameters $\Theta_\alpha, \Theta_\beta $:
\begin{equation}
        F_{\alpha\beta} = - \left\langle \frac{\partial^2 \ln L}{\partial\Theta_\alpha \partial \Theta_\beta} \right\rangle \;.
\end{equation}
From this matrix, the marginal error on parameter $\Theta_\alpha$ in particular can be extracted
as $\sqrt{(F^{-1})_{\alpha \alpha}}$ , and the error on $\Theta_\alpha$, all other parameters being fixed, is bounded by $(F_{\alpha \alpha})^{-1/2}$.

The Fisher matrix may be computed from the covariance matrix of the observable and its derivatives as
\begin{equation}
        F_{\alpha \beta} = \frac{1}{2} Tr[C^{-1} C_{,\alpha} C^{-1} C_{,\beta}] \;.
\end{equation}

\subsubsection{Implementing the tomographic Fisher matrix}

For the tomographic spectra $C_\ell^{(i j)}$, we computed the covariances between spectra under the Gaussian approximation following the approach of \citet{Hu:2004} and \citet{Joachimi:2009}. Denoting by $\Delta C_\ell^{(i j)}$ the difference between the ensemble average of the spectrum and its estimator, the tomographic power spectra covariance is defined as
\begin{align}
\mathrm{Cov}^{(ijkl)}_\ell &\equiv \left\langle \Delta C_\ell^{(i j)} \Delta C_\ell^{(k l)}\right\rangle \;, \\
        &=  \frac{\delta_{\ell \ell^\prime}}{f_{sky} (2 \ell + 1)} \left[ \bar{C}_\ell^{(ik)} \bar{C}_\ell^{(jl)} +\bar{C}_\ell^{(il)}\bar{C}_\ell^{(jk)} \right] \;,
\end{align}
where $f_{sky}$ accounts for partial coverage of the sky and $\bar{C}_\ell^{(ij)}$ is the tomographic power spectrum including shot noise defined in Eq. \eqref{eq:tomopowspec}. The expression of the tomographic Fisher matrix becomes
\begin{equation}
F_{\alpha\beta}^{tomo} = \sum\limits_{(ij), (kl)} \sum\limits_{\ell}^{\ell_{\max}(ijkl)} \frac{\partial C_\ell^{(ij)}}{\partial \Theta_\alpha} {\mathrm{Cov}^{-1}}_\ell^{(ij kl)} \frac{\partial C_\ell^{(kl)}}{\partial \Theta_\beta} \;,
\end{equation} 
where the sum over $(ij), (kl)$ indices loops over all $N_{zbins} (N_{zbins} +1)/2$ combinations of bins, and $\ell_{max}$ is a cut in multipole. The aim of this cut is to restrict the Fisher matrix to linear scales. Several strategies are possible to define $\ell_{\max}$; we describe the one adopted in this work in Sect.~\ref{sec:LinearScales}.

For the binning strategy, we chose to use equal galaxy density bins with no overlap. This choice led to bins with irregular widths, but constant shot noise.

\subsubsection{Implementing the SFB Fisher matrix}

The Fisher matrix for the 3D SFB spectra was computed using the non-diagonal covariance matrix obtained by discretising wavenumbers $k$ under the boundary condition $n(r_{\max}) = 0$ as explained in Sect. \ref{sec:3DSFB}. Details of computing the non-diagonal covariance matrix are given in Appendix~\ref{sec:SFBCov}. In the absence of angular mask, Eq.~\ref{eq:SFBpowspec} shows that the SFB coefficients are uncorrelated between different angular multipoles $\ell$. Therefore, the Fisher matrix for the SFB spectra takes the following form:
\begin{equation}
F^{SFB}_{\alpha \beta} = f_{sky} \sum\limits_{\ell} \frac{(2 \ell + 1 )}{2} \mathrm{Tr}\left[\widehat{C}_\ell^{-1} \frac{\partial \widehat{C}_\ell}{\partial \Theta_\alpha}  \widehat{C}_\ell^{-1} \frac{\partial \widehat{C}_\ell}{\partial \Theta_\beta} \right] \;,
\end{equation}
where the matrices $\widehat{C}_\ell$ are defined as
\begin{equation}
        \widehat{C}_\ell = \left[\begin{matrix} \overline{C}_\ell(0,0) & \overline{C}_\ell(0,1) & \ldots & \overline{C}_\ell(0,n_{\max}^{\ell}) \\
                  \overline{C}_\ell(1,0) & \overline{C}_\ell(1,1) & \ldots & \overline{C}_\ell(1,n_{\max}^{\ell}) \\
                   \vdots& \vdots &\ddots &\vdots\\
                   \overline{C}_\ell(n_{\max}^\ell,0) & \overline{C}_\ell(n_{\max}^\ell,1) & \ldots & \overline{C}_\ell(n_{\max}^{\ell},n_{\max}^{\ell})
        \end{matrix}\right] \;,
\end{equation}
with $\overline{C}_\ell(n,p) = C_\ell(k_{\ell n}, k_{\ell p}) + N_\ell(k_{\ell n}, k_{\ell p})$. The size of each of this matrix $\widehat{C}_\ell$ is $n_{\max}^{\ell} \times n_{\max}^{\ell}$ , where $n_{\max}^\ell$ defines the maximum wavenumber included in the Fisher analysis for each multipole $\ell$. This allows us to restrict the analysis to linear scales. Again, different strategies can be adopted to define this cut in wavenumber; they are described in Sect.~\ref{sec:LinearScales}.

\subsubsection{Fisher analysis baseline}

To conduct this study, we adopted as a fiducial model a `Vanilla' concordance flat cosmology with $h = 0.7$, $\Omega_b=0.045$, $\Omega_m = 0.25$, $\Omega_\Lambda = 0.75$, $\Omega_b = 0.045$, $w_0 = -0.95$, $w_a = 0$, $n_s = 1$, $\tau = 0.09$, and $\sigma_8 = 0.8$. We adopted the standard parametrisation for the dark energy equation of state \citep{Chevallier:2001},\begin{equation}
        w(a) = w_0 + w_a (1 - a) \;.
\end{equation}
This fiducial cosmology was also used to compute the $\tilde{r} = r_{\mathrm{fid}}(z)$. In this model, we computed the linear matter power spectra, including baryonic oscillations, using the fitting formula of \cite{EisensteinHu:1998}. We performed our Fisher analysis on the following parameters $\Theta = (h, \Omega_m, w_0, w_a, \Omega_b, n_s, \sigma_8)$ under the constraint of a flat cosmology.

For our baseline analysis we considered a spectroscopic survey with a very small redshift uncertainty $\sigma_z = 0.003(1+z)$ and a Smail-type galaxy distribution $p(z)$ \citep{Smail:1994},
\begin{equation}
        p(z) \propto z^{2} e^{ - \left( \frac{z}{0.708} \right)^{1.5}} \;,
\end{equation}
which corresponds to a median redshift of $z_{\mathrm{med}} = 1,$ and we used a mean number density of galaxies of $\bar{n} = 0.9$ gal. arcmin$^{-2}$. To account for partial coverage of the sky, we scaled the Fisher information by $f_{sky} = 0.3636,$ which corresponds to a survey size of 15,000 square degrees. This setting was chosen to correspond to the specification of the stage-IV Euclid spectroscopic survey \citep{Redbook}.

Finally, we adopted a redshift dependent fiducial galaxy bias of the form
\begin{equation}
        b(z, k) = \sqrt{1 + z} \;,
        \label{eq:bias}
\end{equation} as in \cite{Rassat:2008bao}.
In Sect.~\ref{sec:NuisanceParameters} we describe how we accounted for our lack of knowledge on the actual galaxy bias by parametrising this relation through nuisance parameters.

\subsubsection{Restriction to linear scales}
\label{sec:LinearScales}

The constraints we aim to extract from a galaxy survey result from the information contained in the matter power spectrum. However, since the galaxies are only biased tracers of the actual underlying matter density, our knowledge of the matter power spectrum is limited by our understanding of the bias. This bias becomes more uncertain on small non-linear scales. Assuming an optimistic knowledge of the bias could result in overestimated or cosmologically biased constraints. Hence, following previous galaxy clustering studies \citep[e.g.,][]{Rassat:2007KRL,Rassat:2008bao,Joachimi:2009}, we completely discarded the mildly to non-linear scales and express our uncertainty of the bias on large scales by using nuisance parameters in the next section.

As the aim of this work is to compare the constraining power of two different approaches to galaxy clustering analysis, it is important to apply the exclusion of non-linear scales to the two methods in a coherent way to avoid biasing our results towards the method with the less conservative cut. Following the approach taken in \citet{Joachimi:2009}, which was based on results from \citet{Rassat:2008bao}, we aim to only retain linear scales through the following redshift-dependent cut in wavenumber $k_{\mathrm{lin}}^{\max}$:
\begin{equation}
        k_{\mathrm{lin}}^{\max}(z) \approx \min[0.132 z, \quad 0.25] \quad h \mathrm{Mpc}^{-1} \;.
\end{equation}
This formula is a linear fit to the non-linearity scale in Fig. 2 of \citet{Rassat:2008bao}, which was computed as a function of redshift by selecting scales that fulfil  $\sigma(R) < 0.20$ and $k_{\max} < 0.25 h$Mpc$^{-1}$, where $\sigma(R)$ corresponds to the amplitude of fluctuations at $R ~h$Mpc$^{-1}$. However, it provides a conservative cut at lower redshift (below z=0.5). Since the purpose of this work is to compare two methodologies given the same framework and set of assumptions, we used this model for the sake of simplicity. An accurate computation of the non-linear scale could be used just as well, but this is not expected to change the conclusions of the comparative analysis.

Because we computed the tomographic power spectra within the Limber approximation, we related wavenumbers $k$ to angular modes $\ell$ through $k = \frac{\ell + 1/2}{r}$. As a result, the non-linear scale cut translates into multipoles $\ell$ for redshift bin $(i)$ as
\begin{equation}
        \ell_{max}^{(i)} = k_{\mathrm{lin}}^{\max}(z_{\mathrm{med}}^{(i)}) r( z_{\mathrm{min}}^{(i)})  \;.
\end{equation}
This cut allows us to reject all the multipoles for a given bin $(i)$ that is affected by scales above $k_{\mathrm{lin}}^{\max}(z_{\mathrm{med}}^{(i)})$. When computing the correlation function between two different bins $(i), (j)$ we applied the most conservative cut: $\ell_{max}^{(i j)} = \min(\ell_{max}^{(i)},  \ell_{max}^{(j)})$.

\begin{figure}[t]
        \includegraphics[width=\columnwidth]{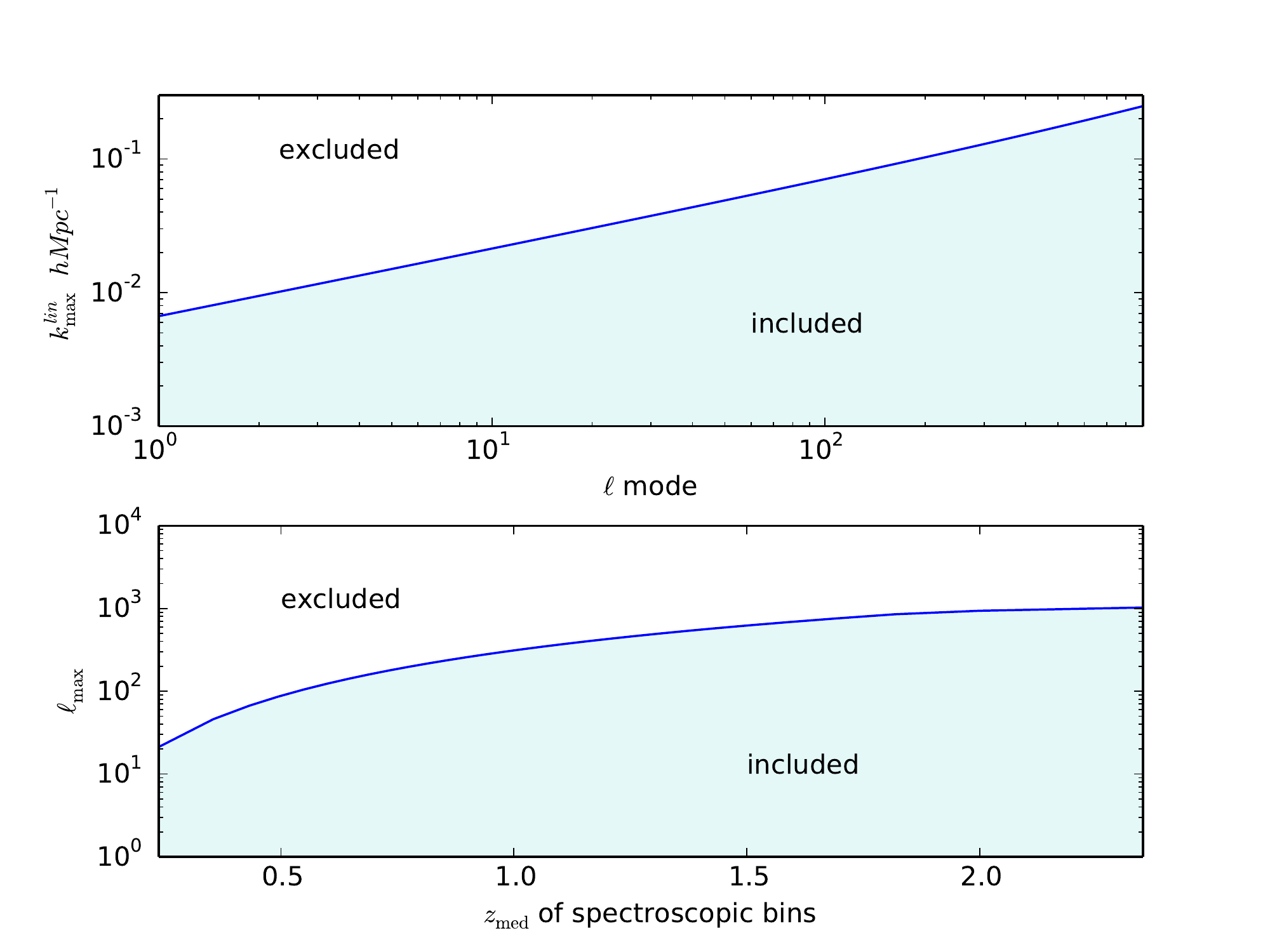} 
        \caption{Top: Linear-scale limit in $k$ for the SFB power spectra as a function of angular modes. Bottom: Linear-scale limit $\ell_{\max}$ for the tomographic power spectra $C^{(ij)}_\ell$ as a function of the lowest median redshift of the spectroscopic bins $i$ and $j$. The regions above the lines are excluded from the Fisher analysis.}
        \label{fig:linear}
\end{figure}

In the SFB framework, applying a corresponding wavenumber cut leads to an $\ell$ dependent maximum number of discrete wavenumbers $k_{\ell n}$, noted $n_{\max}^{\ell}$, which can be obtained as the solution of the equation
\begin{equation}
        k_{\ell n_{\max}^{\ell}} r \left( \frac{ 0.132}{k_{\ell n_{\max}^{\ell}}} \right) = \ell \;,
\end{equation}
under the constraint $k_{\ell n_{\max}^{\ell}} \leq  0.25 \quad h \mathrm{Mpc}^{-1}$. Both cuts are illustrated in Fig.~\ref{fig:linear}.

Thanks to this prescription, the same scales are excluded from the tomographic and SFB analysis. This point is the main difference between our work and the analysis performed in \citet{Nicola:2014}, where the exclusion of non-linear scales is not coherent between the two methodologies. In their work, the angular power spectra are truncated at $\ell_{max} = 50$ for all redshifts, whereas as shown in Fig.~\ref{fig:linear}, $\ell_{max}$ should be a function of the median redshifts of the tomographic bins to take into account the time evolution of the non-linear scale as well as the physical size of angular modes as a function of redshift. Similarly, in the SFB analysis performed in their work, a fixed cut at $k_{\max}=0.20 ~ h \mathrm{Mpc}^{-1}$ was applied, which not only ignores the interplay between angular modes and $z$ illustrated by Fig.~\ref{fig:linear}, but is also incoherent with the cut applied in the tomographic analysis.

\subsubsection{Nuisance parameters}
\label{sec:NuisanceParameters}

As mentioned in the previous section, restricting the study to linear scales avoids the high uncertainty on the bias that arises in the non-linear regime. Nevertheless, we also wish to express our uncertainty on the bias even on linear scales. Following the approach of \citet{Bridle:2007}, \citet{Joachimi:2009}, and \citet{DK:2012}, we parametrised the bias in redshift and scale using a grid of nuisance parameters such that the galaxy bias becomes\begin{equation}
         b(k, z) = A Q(k, z) b_0(k,z) \;,
\end{equation}
where $b_0$ is our fiducial bias relation \eqref{eq:bias}, $A$ is an overall amplitude and $Q(k, z)$ encodes perturbations around the fiducial bias and is defined in terms of an $N_z \times N_k$ grid of parameters $B_{i j}$:
\begin{multline}
 \ln Q(k, z) = K_{i}(k) Z_{j}(z) B_{i j} + [1 - K_{i}(k) ] Z_{j} B_{(i+1) j} \\
 + K_{i} [ 1 - Z_{j}(z)] B_{i (j+1)} + [1 - K_{i}(k) ] [ 1 - Z_{j}(z)] B_{(i+1)(j+1)} \;,
\end{multline}
for $k_i \leq k \leq k_{i+1}$ and $z_j < z \leq z_{j+1}$, where the coefficients $Z_j$ and $K_i$ are expressed as
\begin{align}
K_i(k) =& \frac{\ln(k) - \ln( k_i)}{ \ln(k_{i+1}) - \ln(k_i)} \;, \\
Z_j(k) =& \frac{\ln( 1 + z) - \ln(1 + z_j)}{\ln(1 + z_{j+1}) - \ln(1 + z_j)} \;.
\end{align}
The $k_i$ and $z_j$ fix the nodes of the grid and are spaced logarithmically in the intervals $k \in [10^{-4}, 1.0]$ and $z \in [0, 5]$ such that $k_0 = k_{\min}$, $k_{N_k+1} = k_{\max}$ and $z_0 = z_{\min}$, $z_{N_z + 1} = z_{\max}$. The Fisher matrices are then obtained by marginalising over these $N_k \times N_z +1$ nuisance parameters $\left( A, B_{0 0}, B_{0 1}, B_{1 0}, \ldots , B_{N_k N_z} \right)$. 

\subsection{Figures of merit}
\label{sec:FoM}

Throughout the rest of this work we compare the constraining power of the tomographic and SFB methods by evaluating their respective figures of merit (FoM). We consider two FoMs, first the total figure of merit $\FoMToT$ defined according to \cite{Joachimi:2009} as
\begin{equation}
        \FoMToT = \ln\left( \frac{1}{\det(F^{-1}) } \right) \;,
\end{equation}
and second, the dark energy figure of merit recommended by the report of the Dark Energy Task Force \citep{DETF},
\begin{equation}
        \FoMDE = \frac{1}{\sqrt{\det( F^{-1} )_{w_0 w_a}}} \;.
\end{equation}
The DETF FoM was designed to measure the strength of a given future survey or probe in constraining cosmological parameters related to the nature of dark energy, such that a large FoM$_{\rm DETF}$ value meant a high constraining power on $w_0$ and $w_a$. The total FoM (FoM$_{\rm TOT}$) was designed to encompass the strength of a future survey or probe in constraining several parameters across different sectors of cosmology, such as the nature of dark matter and dark energy and initial conditions. A high value of FoM$_{\rm TOT}$ therefore means a good constraining power across all cosmological sectors. The parameter is taken as an $\ln$ value, since we consider this number for seven cosmological parameters, and the FoM$_{\rm TOT}$ value would grow very quickly otherwise. 
\section{Results: SFB vs. tomographic analysis}
\label{sec:results}

\subsection{Comparison of SFB and tomographic analysis in the absence of systematics}
\label{results:3d2d}

\begin{figure}[t]
   \centering
   \includegraphics[width=\columnwidth]{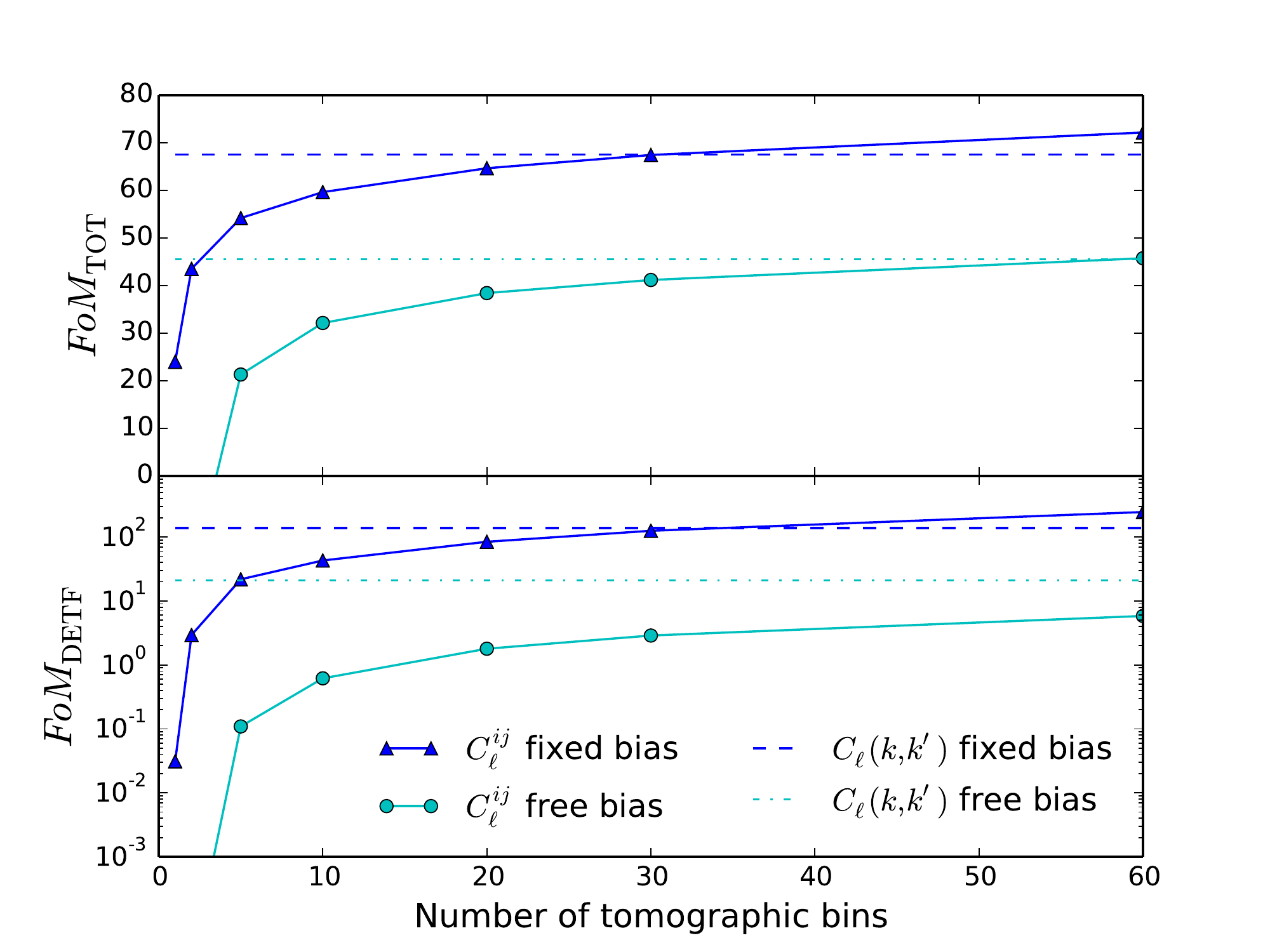} 
   \caption{Comparison of the total ${\rm FoM}_{\mathrm{TOT}}$ (top) and dark energy ${\rm FoM}_{\mathrm{DETF}}$ (bottom) figures of merit for the 3D Fourier-Bessel analysis (horizontal dashed lines) vs. tomographic analysis (solid lines) as a function of number of redshift bins. The upper lines (dashed and triangle) result from assuming a fixed bias, the lower lines (dotted and circle) are obtained when assuming a grid of $5 \times 5$ nuisance parameters in scale and redshift described in Sect.~\ref{sec:NuisanceParameters}.}
   \label{fig:3dvs2d}
\end{figure}

Here, we compare the relative constraining power of the tomographic and SFB analysis of galaxy clustering presented in Sect.\ref{sec:theory} using the Fisher matrix formalism and the fiducial cosmology and survey baseline described in Sect.~\ref{sec:Forecasting}. We investigate first the impact of the number of redshift bins and whether the same constraints can be recovered from the two different analysis. Figure~\ref{fig:3dvs2d} shows the FoMs obtained using both methods as a function of number of tomographic spectroscopic bins when assuming perfect knowledge of the bias (in dark blue).

\begin{figure*}[!ht]
   \centering
   \includegraphics[width=\textwidth]{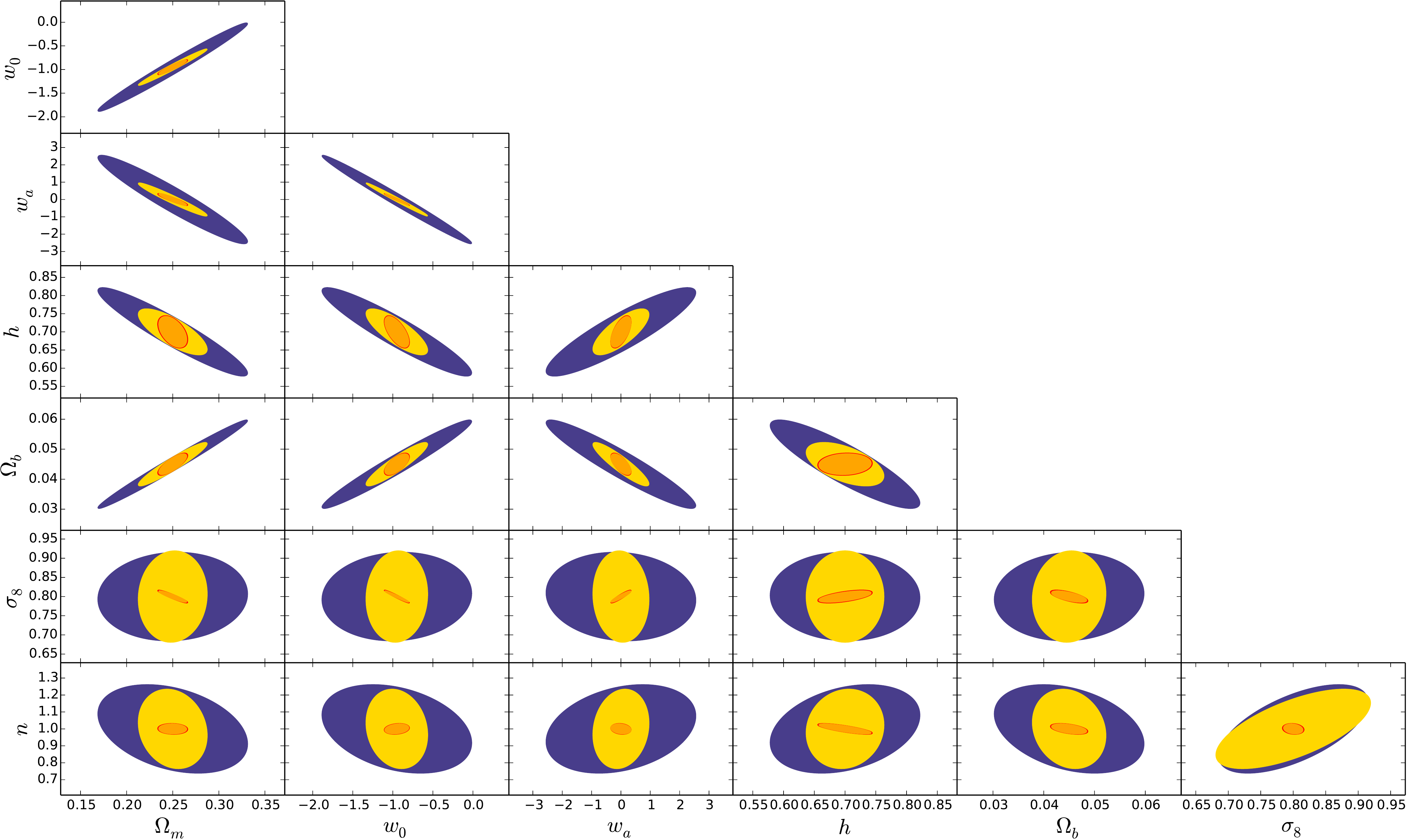} 
   \caption{1$\sigma$ contours for all pairs of cosmological parameters for the SFB analysis and the tomographic analysis for 30 bins with and without nuisance parameters. Inner orange and red contours (almost indistinguishable) result from the SFB and tomographic analysis when assuming a fixed bias. Outer yellow and purple contours are obtained from the SFB and tomographic analysis when using a $5 \times 5$ nuisance parameter grid in scale and redshift for the bias.}
   \label{fig:constraints}
\end{figure*}

As expected, the two figures of merit for the tomographic analysis increase with the number of redshift bins and eventually reach the performance of the SFB analysis for 30 redshift bins. Not only do the two methodologies yield equivalent figures of merit for this number of bins, but the $1\sigma$ contours for all cosmological parameters are extremely similar, both in size of the ellipse and for the direction of the degeneracies. Figure~\ref{fig:constraints} shows the $1\sigma$ contours on all pairs of cosmological parameters considered for the two analysis techniques using 30 tomographic bins with and without nuisance parameters for the bias. For the fixed bias, the contours obtained by the tomographic analysis are plotted in red and are almost indistinguishable from the contours for the SFB analysis, which are depicted in orange. 

We conclude that exactly the same information is extracted from the two methodologies for an appropriate number of redshift bins, 30 in our case.

This result disagrees with the conclusions of \citet{Nicola:2014}, who found that the SFB analysis is weaker than a tomographic analysis and not capable of extracting the same radial information. The difference in these conclusions is probably related to the choice of non-linear prescription. In \cite{Nicola:2014}, the tomographic SHT analysis was limited to a fixed $\ell_{\max}$ for all bins, while the SFB analysis was restricted to a fixed $k_{\max}$ for all multipoles. However, we show in Fig.~\ref{fig:linear} that to apply equivalent cuts for the tomographic and SFB analysis, redshift-dependent $\ell_{\max}(z_{med})$ and $\ell$-dependent $k_{\max}(\ell)$ cuts need to be used.

\bigskip

Additionally, Fig.~\ref{fig:3dvs2d} shows that when the number of bins is increased, the tomographic analysis eventually surpasses the SFB analysis. This behaviour is expected, because when the width of the redshift bins reaches the non-linearity scale, the tomographic analysis probes more modes than a 3D analysis \citep{Asorey:2012,Clzz}. Indeed, only non-linear angular scales are excluded from the tomographic analysis, but for very thin redshift bins, small radial scales are being probed that are potentially beyond the non-linear cut-off. \citet{Asorey:2012} found that a tomographic analysis with a bin width of $\Delta r \simeq 0.8 \frac{2 \pi}{k_{\max}}$ was equivalent to a 3D power spectrum analysis including scales up to $k_{\max}$. We found that the tomographic analysis recovers the information from the 3D analysis for about 30 redshift bins. If one expects the two methodologies to give similar results for $\Delta r \simeq \frac{2\pi}{k_{max}}$ , then one would expect a larger number of tomographic bins to be necessary. Here, our 30 bins correspond to a minimum bin width $\Delta r \simeq 0.55 \frac{2\pi}{k_{max}}$ , which is not as close to the non-linearity scale as the results from \citet{Asorey:2012}, but remains of the same order of magnitude.

However, we stress that such a direct comparison is subject to several factors that complicate the interpretation. Firstly, the tomographic spectra are computed within the Limber approximation, which may not be accurate for a large number of thin bins. A recent study of the effect of the Limber approximation for a
spectroscopic survey can be found in \citet{Eriksen:2014}. Because we restricted our analysis to large linear scales, we limited the number of tomographic bins to 30 in the rest of the analysis, which corresponds to redshift widths between $\Delta_z=0.1$ and $\Delta_z=0.05$. In this case, according to \citet{Eriksen:2014}, the error of approximation remains limited (below 15\% for most bins). Therefore, we do not expect the full computation to significantly alter the results of the comparative study lead in this work. Nevertheless, this point should be kept in mind and deserves a thorough analysis, which we will include in future work. We also stress that although care has been taken to apply similar non-linear cuts, they are not strictly equivalent, and different strategies to restrict angular modes in the tomographic analysis would affect the results.

Therefore, we consider that for a fixed bias, both analysis methodologies recover the same information for 30 tomographic bins, which corresponds to a minimum bin width of the order of the non-linearity scale. We also acknowledge that the exact number of bins is likely to change for different binning strategy, computation techniques of angular power spectra, restrictions of non-linear scales and with the inclusion of additional effects such as redshift space distortions or relativistic effects. A thorough study of all these effects will be addressed in a future paper.

\subsection{Impact of systematics due to galaxy bias}\label{sec:nuisance}

After establishing that the same information can be recovered from both methodologies in the absence of systematics on the bias, we now investigate the impact of an unknown bias. As described in Sect.~\ref{sec:NuisanceParameters}, we include in the analysis an uncertainty on the galaxy bias using a grid of nuisance parameters in scale and redshift. Figure~\ref{fig:3dvs2d} demonstrates how the FoMs for both analysis are degraded when using a free bias parametrised in scale and redshift by a $5\times 5$ nuisance parameter grid (in cyan). Whereas the FoMs were equivalent with 30 tomographic bins in the fixed bias case, the tomographic analysis can no longer recover the same information as the SFB analysis in the free bias case, even with 60 redshift bins. The tomographic analysis is much more sensitive to systematics resulting from the unknown bias than the SFB analysis. 

\begin{figure}[t]
   \centering
   \includegraphics[width=\columnwidth]{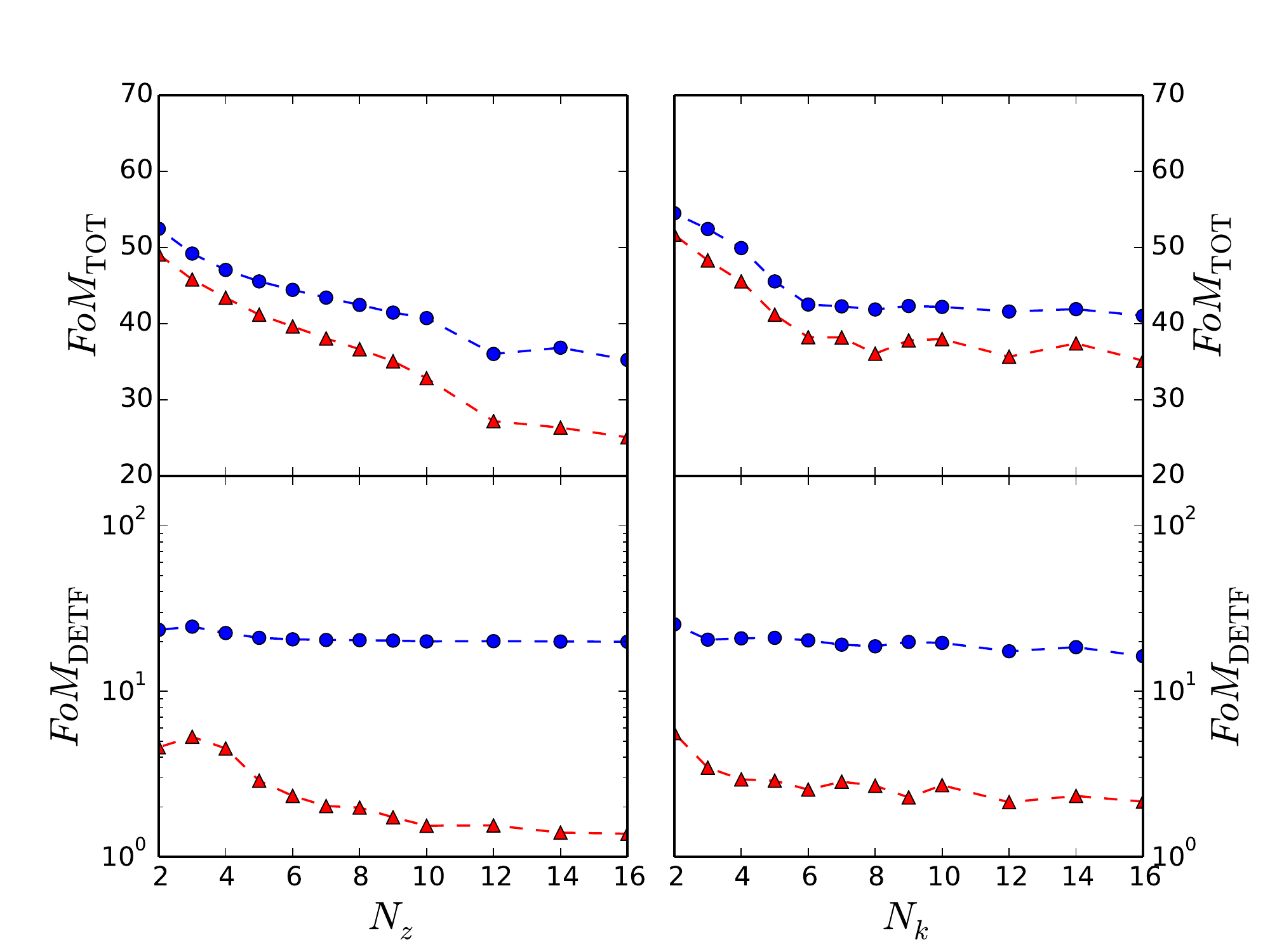} 
   \caption{Total FoM$_{\rm TOT}$ (top) and dark energy FoM$_{\rm DETF}$ (bottom) figures of merit as a function of the number of nuisance parameters in redshift (left) and scale (right), for a tomographic with 30 bins (red triangle) and an SFB (blue dot) analysis. When varying the number of nuisance parameters in scale or redshift, the other number of parameters is kept fixed at 5.}
   \label{fig:nuisance}
\end{figure}

We investigated the effect of the number of nuisance parameters in scale and redshift $N_k \times N_z$ on the FoMs for the tomographic (red triangle) and SFB (blue dot) analysis in Fig.~\ref{fig:nuisance}. We varied $N_k$ and $N_z$ independently while keeping the other parameter fixed to $5$. When the number of nuisance parameters increases, the constraints from both analyses decreases, although the FoMs from the tomographic analysis degrade faster than for the SFB analysis. 

Although the FoMs reach a plateau at about $N_z=12$ and $N_k=6$, these numbers would correspond to a very conservative model of the galaxy bias and therefore are probably unrealistic. Indeed, the evolution of galaxy bias should be smooth on large scales, which prompts us to limit the fiducial parameter grid used in this section to $N_z=5$ and $N_k=5$. Since the trends in FoMs do not change with the number of nuisance parameters, a more complex grid (increasing either $N_z$ or $N_k$) would not change the conclusions on the relative strength of the two methodologies investigated here (the SFB FoM remains higher for any choice of nuisance parameters).

The effect of the free bias on the $1\sigma$ contours on cosmological parameters is shown in Fig.~\ref{fig:constraints}, where the purple and yellow contours are computed from the 30-bin tomographic analysis and the SFB analysis. Interestingly, the constraints on $\sigma_8$ and $n_s$ are affected in the same way for the two methodologies by the inclusion of nuisance parameters; the contours are almost equivalent for $(n_s,\sigma_8)$ with or without nuisance parameters. In contrast, all other parameters are much more degraded by the including nuisance parameters in the case of the tomographic analysis compared to the SFB analysis. This is particularly true for the dark energy parameters $w_0$ and $w_a$.

These results agree with \cite{Asorey:2012}, who noted that the tomographic constraints degrade faster than a 3D power spectrum analysis when a single nuisance parameter on the amplitude of the bias was included. We find a similar behaviour with a more flexible parameterisation of the bias and for the 3D SFB analysis.

Furthermore, these results highlight the well-known sensitivity of galaxy clustering studies to the galaxy bias, which is one of its most important systematics. Although other approaches such as the measurement of the BAO scale are less sensitive to the galaxy bias, this results in the usual trade-off between systematics and statistical constraining power, so that BAO studies alone (i.e. using only BAO scale measurement) only provide conservative constraints without relying on external priors \citep{Rassat:2008bao}.

\subsection{Optimisation of a stage-IV survey}
Since we have shown in Sect. \ref{sec:nuisance} that the 3D SFB and tomographic methods depend differently on nuisance parameters, we are interested in investigating whether there are other differences in using one method or the other to plan for future wide-field surveys. 

\begin{figure}[t]
   \centering
   \includegraphics[width=\columnwidth]{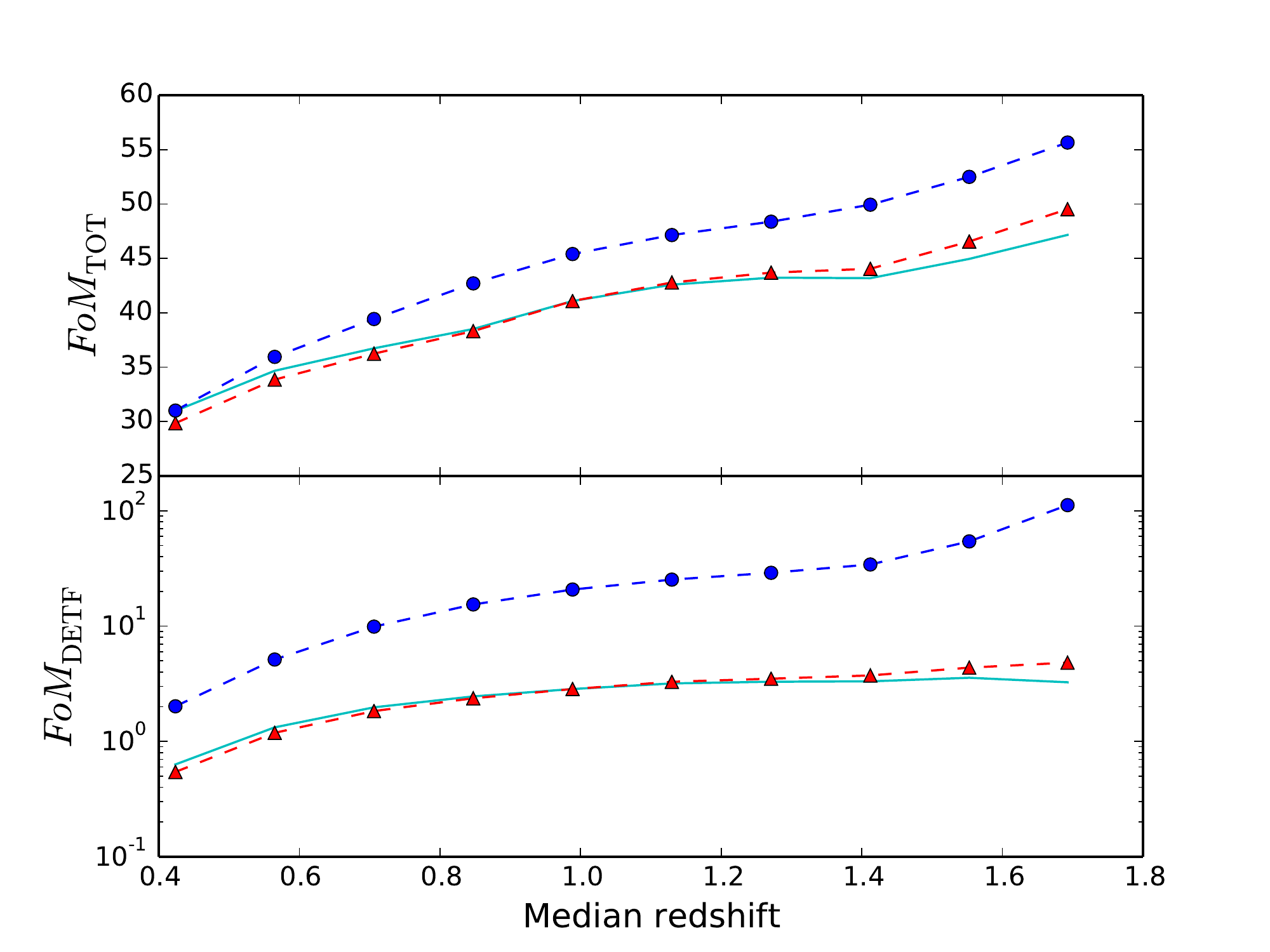} 
   \caption{Total ${\rm FoM}_{\mathrm{TOT}}$ (top) and dark energy ${\rm FoM}_{\mathrm{DETF}}$ (bottom) figures of merit as a function of median redshift of a stage-IV spectroscopic survey using a tomographic analysis (red triangles) and an SFB analysis (blue dots). The cyan line shows the figures of merit for the tomographic analysis when the number of bins is not adapted to the depth of the survey and kept at 30 bins. For the red-dashed lines, the number of bins has been adapted to each median redshift. In all cases, a grid of $5 \times 5$ nuisance parameters in scale and redshift is used to parametrise the galaxy bias.}
   \label{fig:opt}
\end{figure}

In this section we investigate the influence of the median redshift on the constraining power of a stage-IV spectroscopic survey using the two techniques. To perform this comparison, we used the same  $5 \times 5$ nuisance parameter grid for the bias as in the previous section. We also adapted the number of tomographic bins to the median redshift of the survey to preserve the equivalence between tomographic and SFB constraints in the absence of systematics found in Sect.~\ref{results:3d2d}. The smallest radial scales probed by a tomographic analysis depend on the depth of the survey and on the number of bins. Therefore, to remain coherent for different median redshifts with the SFB analysis, the number of bins needs to be adjusted to the median redshift.
We find that for a median redshift of $z_{med} \simeq 0.4$, the number of bins of the tomographic analysis should be $N=26$ and for $z_{med} \simeq 1.7$ this number increases to $N=42$.
To illustrate this point, we plot in Fig.~\ref{fig:opt} the FoMs
as red triangles as a function of the median redshift using this adapted number of bins. The cyan line shows the evolution of the FoMs when keeping the number of bins fixed at $N=30$. Since this number of 30 tomographic bins was chosen in the previous section based on our fiducial survey with a median redshift of 1, we see that the red and cyan curves cross at $z_{med}=1$. However, using 30 bins below $z_{med}=1$ means probing smaller radial scales, which are beyond the scales probed by the SFB analysis, and this increases the FoMs. In contrast, above $z_{med}=1$, this means using wider tomographic bins and thus probing larger radial scales, which lowers the FoMs compared to when the number of bins is adapted.

We also plot in Fig.~\ref{fig:opt} the 3D SFB FoMs as a function of the median redshift of the survey, using blue circles. This curve should be compared to the red triangles showing the FoMs for the tomographic analysis where we have adapted the number of tomographic bins based on the median redshift, as described above.
The two techniques exhibit a similar scaling with the median redshift of the survey, although the SFB constraints are consistently better than the tomographic constraints. Interestingly, for median redshifts above $z_{med}=1.4$, the SFB dark energy FoM exhibits a better scaling than the tomographic one.

In conclusion, in the presence of galaxy bias systematics, any desired FOM level can be reached for shallower surveys if a 3D SFB analysis is performed. Furthermore, increasing the depth of the survey is more profitable in terms of FOM$_{\rm DETF}$ for the 3D SFB analysis because the tomographic method reaches
a plateau somewhat after $z=1$, whereas the 3D method continues to increase significantly up to $z=1.8$, potentially pushing the optimisations towards higher median redshifts.

\section{Conclusion}
\label{sec:ccl}

We have compared two different approaches to the three-dimensional analysis of galaxy clustering in the context of wide and deep future spectroscopic galaxy surveys. Based on the Fisher matrix analysis, we have compared the tomographic spherical harmonics and spherical Fourier-Bessel (SFB) methodologies in terms of figures of merit and cosmological parameter constraints.

We focused on the seven common parameters that are currently used in wide-field survey optimisation and planning: $\vec{\theta}=\{\Omega_m, h, w_0, w_a, \sigma_8, \Omega_b, n_s\}$, while putting forward a coherent and realistic approach regarding the exclusion of non-linear scales for both the 2D and 3D methods. In addition, we investigated for the first time how tomographic and 3D SFB methods are affected by nuisance parameters related to the galaxy bias, which we allowed to be both redshift- and scale-dependent. 

In the absence of systematics, for an appropriate number of tomographic bins the two methodologies are equivalent and are able to recover the exact same constraints - both in value and in direction of degeneracy between different parameters. Increasing the number of redshift bins further leads to stronger constraints for the tomographic analysis, as seen by \citet{Asorey:2012}, \citet{Clzz}, and \citet{Nicola:2014}. Nevertheless, this effect could result from including radial
scales in the tomographic analysis that are beyond the non-linear cut-off applied to the SFB analysis, and should be investigated further.

On the other hand, when we included unavoidable systematics due to the galaxy bias through a grid of nuisance parameters in scale and redshift, we found that the SFB analysis is more robust than the tomographic analysis, whose constraints suffer more from including nuisance parameters. As a result, we found that when we optimised the median redshift of a stage-IV type spectroscopic galaxy survey, a given level of accuracy can be achieved for shallower surveys if a 3D SFB analysis is performed. Moreover, the scaling of the dark energy figure of merit with median redshift is better for the 3D SFB analysis in the presence of systematics, which means that a given increase of the survey depth yields more information using an SFB analysis than a tomographic analysis.

Our results suggest that an SFB analysis is preferable to a tomographic analysis for realistic future spectroscopic wide-field surveys where the galaxy bias can be both redshift- and scale-dependent, and is unknown. These conclusions should be investigated in more
detail, for example regarding the potential effect of the exact computation of angular power spectra, binning strategy, and including RSD.

In the spirit of reproducible research, the Python package \textbf{CosmicPy} developed to produce all the results presented in this work is freely available at\\ \\
{\centerline{\url{http://cosmicpy.github.io} .}}\\ \\
This package allows for simple and interactive computation of tomographic and 3D SFB power spectra as well as Fisher matrices while relying on a fast C++ implementation of Fourier-Bessel related computations.

\begin{acknowledgements}
The authors thank Chieh-An Lin and Martin Kilbinger for useful comments and discussions. This research is in part supported by the European Research Council grant SparseAstro (ERC-228261) and by the Swiss National Science Foundation (SNSF).
\end{acknowledgements}

\bibliographystyle{aa}
\bibliography{references13}

\newpage
\begin{appendix}

\section{ CosmicPy package}
\label{sec:CosmicPy}

\begin{listing}
\begin{Verbatim}[commandchars=\\\{\},codes={\catcode`\$=3\catcode`\^=7\catcode`\_=8},frame=lines]
\PY{o}{\PYZgt{}\PYZgt{}}\PY{o}{\PYZgt{}} \PY{k+kn}{from} \PY{n+nn}{cosmicpy} \PY{k+kn}{import} \PY{o}{*}

\PY{c}{\PYZsh{} Create a standard cosmology}
\PY{o}{\PYZgt{}\PYZgt{}}\PY{o}{\PYZgt{}} \PY{n}{cosmo} \PY{o}{=} \PY{n}{cosmology}\PY{p}{(}\PY{p}{)}
\PY{o}{\PYZgt{}\PYZgt{}}\PY{o}{\PYZgt{}} \PY{k}{print} \PY{n}{cosmo}
\PY{n}{FLRW} \PY{n}{Cosmology} \PY{k}{with} \PY{n}{the} \PY{n}{following} \PY{n}{parameters}\PY{p}{:} 
    \PY{n}{h}\PY{p}{:}        \PY{l+m+mf}{0.7} 
    \PY{n}{Omega\PYZus{}b}\PY{p}{:}  \PY{l+m+mf}{0.045} 
    \PY{n}{Omega\PYZus{}m}\PY{p}{:}  \PY{l+m+mf}{0.25} 
    \PY{n}{Omega\PYZus{}de}\PY{p}{:} \PY{l+m+mf}{0.75} 
    \PY{n}{w0}\PY{p}{:}       \PY{o}{\PYZhy{}}\PY{l+m+mf}{0.95} 
    \PY{n}{wa}\PY{p}{:}       \PY{l+m+mf}{0.0} 
    \PY{n}{n}\PY{p}{:}        \PY{l+m+mf}{1.0} 
    \PY{n}{tau}\PY{p}{:}      \PY{l+m+mf}{0.09} 
    \PY{n}{sigma8}\PY{p}{:}   \PY{l+m+mf}{0.8}

\PY{c}{\PYZsh{} Setup a spectroscopic survey specifying the}
\PY{c}{\PYZsh{} redshift distribution, fsky, redshift}
\PY{c}{\PYZsh{} errors and galaxy density per square arcmin.}
\PY{o}{\PYZgt{}\PYZgt{}}\PY{o}{\PYZgt{}} \PY{n}{surv} \PY{o}{=} \PY{n}{survey}\PY{p}{(}\PY{n}{nzparams}\PY{o}{=}\PY{p}{\PYZob{}}\PY{l+s}{\PYZsq{}}\PY{l+s}{type}\PY{l+s}{\PYZsq{}}\PY{p}{:}\PY{l+s}{\PYZsq{}}\PY{l+s}{smail}\PY{l+s}{\PYZsq{}}\PY{p}{,}
                           \PY{l+s}{\PYZsq{}}\PY{l+s}{a}\PY{l+s}{\PYZsq{}}\PY{p}{:}\PY{l+m+mf}{2.0}\PY{p}{,}
                           \PY{l+s}{\PYZsq{}}\PY{l+s}{b}\PY{l+s}{\PYZsq{}}\PY{p}{:}\PY{l+m+mf}{1.5}\PY{p}{,}
                           \PY{l+s}{\PYZsq{}}\PY{l+s}{z0}\PY{l+s}{\PYZsq{}}\PY{p}{:}\PY{l+m+mf}{0.71}\PY{p}{\PYZcb{}}\PY{p}{,}
                 \PY{n}{fsky}\PY{o}{=}\PY{l+m+mf}{0.3636}\PY{p}{,}
                 \PY{n}{zphot\PYZus{}sig}\PY{o}{=}\PY{l+m+mf}{1e\PYZhy{}3}\PY{p}{,}
                 \PY{n}{ngal}\PY{o}{=}\PY{l+m+mf}{0.9}\PY{p}{,}
                 \PY{n}{nzbins}\PY{o}{=}\PY{l+m+mi}{30}\PY{p}{)}
           
\PY{c}{\PYZsh{} Cosmological parameters to include in the}
\PY{c}{\PYZsh{} Fisher matrix analysis}
\PY{o}{\PYZgt{}\PYZgt{}}\PY{o}{\PYZgt{}} \PY{n}{params} \PY{o}{=} \PY{p}{(}\PY{l+s}{\PYZsq{}}\PY{l+s}{Omega\PYZus{}m}\PY{l+s}{\PYZsq{}}\PY{p}{,} \PY{l+s}{\PYZsq{}}\PY{l+s}{w0}\PY{l+s}{\PYZsq{}}\PY{p}{,} \PY{l+s}{\PYZsq{}}\PY{l+s}{wa}\PY{l+s}{\PYZsq{}}\PY{p}{,} \PY{l+s}{\PYZsq{}}\PY{l+s}{h}\PY{l+s}{\PYZsq{}}\PY{p}{,}
              \PY{l+s}{\PYZsq{}}\PY{l+s}{Omega\PYZus{}b}\PY{l+s}{\PYZsq{}}\PY{p}{,} \PY{l+s}{\PYZsq{}}\PY{l+s}{sigma8}\PY{l+s}{\PYZsq{}}\PY{p}{,} \PY{l+s}{\PYZsq{}}\PY{l+s}{n}\PY{l+s}{\PYZsq{}}\PY{p}{)}

\PY{c}{\PYZsh{} Create a 3D SFB Fisher matrix based on the}
\PY{c}{\PYZsh{} input cosmology and survey.}
\PY{o}{\PYZgt{}\PYZgt{}}\PY{o}{\PYZgt{}} \PY{n}{f3d} \PY{o}{=} \PY{n}{fisher3d}\PY{p}{(}\PY{n}{cosmo}\PY{p}{,} \PY{n}{surv}\PY{p}{,} \PY{n}{params}\PY{p}{)}

\PY{c}{\PYZsh{} Create a Tomographic Fisher matrix based on}
\PY{c}{\PYZsh{} the input cosmology and survey.}
\PY{o}{\PYZgt{}\PYZgt{}}\PY{o}{\PYZgt{}} \PY{n}{ftomo} \PY{o}{=} \PY{n}{fisherTomo}\PY{p}{(}\PY{n}{cosmo}\PY{p}{,} \PY{n}{surv}\PY{p}{,} \PY{n}{params}\PY{p}{,} \PY{l+s}{\PYZsq{}}\PY{l+s}{g}\PY{l+s}{\PYZsq{}}\PY{p}{,}
               \PY{n}{cutNonLinearScales}\PY{o}{=}\PY{l+s}{\PYZsq{}}\PY{l+s}{realistic}\PY{l+s}{\PYZsq{}}\PY{p}{)}\PY{p}{)}

\PY{c}{\PYZsh{} Output the total Figure of Merit}
\PY{o}{\PYZgt{}\PYZgt{}}\PY{o}{\PYZgt{}} \PY{n}{f3d}\PY{o}{.}\PY{n}{FoM}
\PY{l+m+mf}{67.516476905327863}

\PY{o}{\PYZgt{}\PYZgt{}}\PY{o}{\PYZgt{}} \PY{n}{ftomo}\PY{o}{.}\PY{n}{FoM}
\PY{l+m+mf}{67.495159354357327}

\PY{c}{\PYZsh{} Display a corner plot for both fisher matrices}
\PY{o}{\PYZgt{}\PYZgt{}}\PY{o}{\PYZgt{}} \PY{n}{ftomo}\PY{o}{.}\PY{n}{corner\PYZus{}plot}\PY{p}{(}\PY{n}{nstd}\PY{o}{=}\PY{l+m+mi}{1}\PY{p}{)}

\PY{o}{\PYZgt{}\PYZgt{}}\PY{o}{\PYZgt{}} \PY{n}{f3d}\PY{o}{.}\PY{n}{corner\PYZus{}plot}\PY{p}{(}\PY{n}{nstd}\PY{o}{=}\PY{l+m+mi}{1}\PY{p}{,}\PY{n}{color}\PY{o}{=}\PY{l+s}{\PYZsq{}}\PY{l+s}{r}\PY{l+s}{\PYZsq{}}\PY{p}{)}
\end{Verbatim}

\caption{Example of 3D SFB and tomographic Fisher matrix computations using CosmicPy.}
\label{lst:CPy}
\end{listing}
\textbf{CosmicPy} is an interactive Python package that allows for simple cosmological computations. Designed to be modular, well-documented and easily extensible, this package aims to be a convenient tool for forecasting cosmological parameter constraints for different probes and different statistics. Currently, the package includes basic functionalities such as cosmological distances and matter power spectra (based on \citet{EisensteinHu:1998} and \citet{Smith:2003p}), and facilities for computing tomographic (using the Limber approximation) and 3D SFB power spectra for galaxy clustering and the associated Fisher matrices.

Listing~\ref{lst:CPy} illustrates how CosmicPy can be used to easily compute the 3D SFB Fisher matrix, extract the figure of merit, and generate the associated corner plot similar to Fig.~\ref{fig:constraints}.  

The full documentation of the package and a number of tutorials demonstrating how to use the different functionalities and reproduce the results of this paper is provided at the CosmicPy webpage: \url{http://cosmicpy.github.io} .

Although CosmicPy is primarily written in Python for code readability, it also includes a simple interface to C/C++, allowing critical parts of the codes to have a fast C++ implementation as well as enabling existing codes to be easily interfaced with CosmicPy.

Contributions to the package are very welcome and can be in the form of feedback, requests for additional features, documentation, or even code contributions. This is made simple through the GitHub hosting of the project at
{\centerline{\url{https://github.com/cosmicpy/cosmicpy} .}}

\section{Computing the SFB covariance matrix}
\label{sec:SFBCov}

Performing a Fisher analysis requires computing the SFB covariance matrix, and more importantly, computing the inverse of this matrix. This last step can be quite challenging as the covariance of the spherical Fourier-Bessel coefficients is a continuous quantity $C_{\ell}(k,k^\prime)$. Two approaches can be considered to define a covariance matrix in this situation: (i) only using the diagonal covariance $C_\ell(k_i,k_i)$ at discrete points $k_i$ (advocated by \citet{Nicola:2014}), or (ii) binning $C_\ell(k,k^\prime)$ into bins of size $\Delta_k$. However, by neglecting the correlation between neighbouring wavenumbers, the first approach overestimates the information content if the interval between wavenumbers is too small, while the second approach would lose information for bins of increasing size and become numerically challenging to invert for bins too small. Another problem is to select the largest scale $k_{\min}$ to include in the covariance matrix. Indeed, $C_\ell(k,k)$ becomes extremely small and numerically challenging to compute for very small $k,$ but small wavenumbers can still potentially contribute to the Fisher information. A careful study is necessary to select a $k_{\min}$ that does not lose information.

\begin{figure}
        \includegraphics[width=\columnwidth]{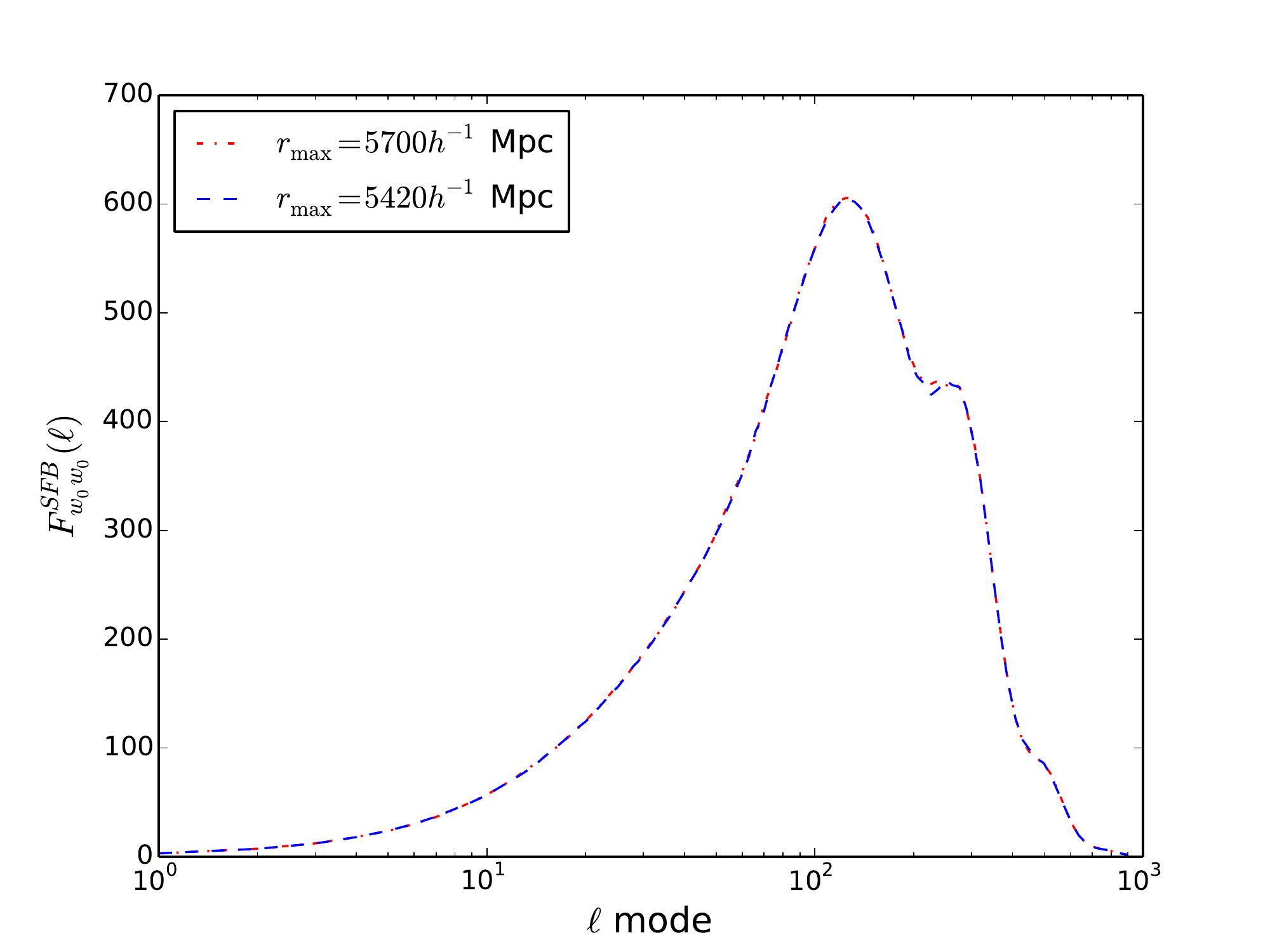}
        \caption{Contribution to the SFB Fisher matrix element $F_{w_0 w_0}^{SFB}$ as a function of angular mode, computed with different values of $r_{\max}$. The excellent agreement between the two curves shows that our computation of the Fisher matrix is robust to our arbitrary choice of $r_{\max}$.}
        \label{fig:FisherElements}
\end{figure}

Instead, using the $k_{l n}$ sampling defined by Eq. \eqref{eq:scale_sampling} naturally introduces a minimum wavenumber and a discrete sampling of scales that preserves all the information. As an added benefit, this approach yields numerically invertible covariance matrices in practice for sensible choices of the boundary condition $r_{\max}$. Indeed, as $k_{l n} = \frac{q_{l n}}{r_{\max}}$, the choice of cut-off radius sets the fineness of the $C_\ell(n,n^\prime)$ matrix and affects its condition number. However, we find that the Fisher information remains largely unaffected by varying $r_{\max}$ above a certain distance because cutting the very end of the galaxy distribution has little effect. In practice, we have arbitrarily set $r_{\max}$ to the comoving distance at which $\phi(r)$ reaches $10^{-5}$ of its maximum value. This choice has proven stable in all situations considered in this work. The robustness of our computation of the Fisher matrix with respect to the choice of $r_{\max}$ is illustrated in Fig.~\ref{fig:FisherElements}, where we show the contributions of each angular mode to the Fisher matrix element $F_{w_0 w_0}^{SFB}$. Our empirical choice for $r_{\max}$ in this case is 5420 $h^{-1}$Mpc, but the results are not affected by increasing $r_{\max}$ to 5700 $h^{-1}$ Mpc
even more.

\section{Deriving the spherical Fourier-Bessel shot noise power spectrum}
\label{sec:ShotNoise}
Here, we derive the expression of the shot noise by discretising the survey in cells that either contain  one or zero galaxies \citep{Peebles:1980}. This method was used in \citet{Heavens:2006} to yield the expression of the shot noise in the case of 3D cosmic shear. We considered a point process defined on small cells $c$, each of which contains $n_c=0$ or $1$ depending on whether the cell contains a galaxy or not:

\begin{equation}
        n(\mathbf{r}) = \frac{1}{V}\sum\limits_c  n_c \delta_c(\mathbf{r}) \;,
\end{equation}
where $\delta_c(\mathbf{r}) = 1$ if $\mathbf{r}$ is within the cell $c$, 0 otherwise, and where $n_c$ fulfils \citep{Peebles:1980}
\begin{equation}
        <n_c> = <n_c^2> = \bar{\rho}_{g}^{obs} \phi(r_c) \Delta_c \;,
\end{equation}
where $\Delta_c$ is the volume of cell $c$ and $\bar{\rho}_{g}^{obs} \phi(r_c)$ is the average number density of galaxies of the survey at distance $r_c$. Furthermore, the cross-term for $c \neq d$ is\begin{equation}
        <n_c n_d> = \left. \bar{\rho}_{g}^{obs} \right.^2 \phi(r_c)\phi(r_d) \Delta_c \Delta_d [1 + \xi(|\mathbf{r_c} - \mathbf{r_d}|) ] \;.
\end{equation}

The SFB expansion of the density field can now be expressed as a sum over small cells $c$:
\begin{equation}
        n_{\ell m}(\mathbf{k}) = \sqrt{\frac{2}{\pi}} \sum\limits_c n_c k j_{\ell}(k r_c)  Y_{\ell m} (\mathbf{\Omega_c}) \;.
\end{equation}
From this expression, we can derive the two-point correlation function of this field:
\begin{align}
        <n_{\ell m}(\mathbf{k}) n_{\ell' m'}(\mathbf{k'})> &= \frac{2}{\pi} \sum_{c,d} <n_c n_d> k k^\prime j_{\ell}(k r_c) j_{\ell'}(k' r_d) \nonumber \\
        & \qquad \times   \overline{Y_\ell^m(\mathbf{\Omega_c})} Y_{\ell'}^{m'} (\mathbf{\Omega_d}) \\
        &= \frac{2}{\pi} \sum_{c = d} \bar{n} \phi(r_c) \Delta_c k k^\prime j_{\ell}(k r_c) j_{\ell'}(k' r_c) \\
        & \qquad \times  \overline{Y_\ell^m(\mathbf{\Omega_c})} Y_{\ell'}^{m'} (\mathbf{\Omega_c})\nonumber \\
        & \quad + \frac{2}{\pi} \sum_{c \neq d} \left. \bar{n} \right.^2 \phi(r_c)\phi(r_d) \nonumber \\ 
        & \qquad \times \Delta_c \Delta_d [1 + \xi_g(|\mathbf{r_c} - \mathbf{r_d}|) ] \nonumber\\
        & \qquad \times k k^\prime j_{\ell}(k r_c) j_{\ell'}(k' r_d) \overline{Y_\ell^m(\mathbf{\Omega_c})} Y_{\ell'}^{m'}(\mathbf{\Omega_d})\nonumber \;.
\end{align}
In the last equation, the first term for $c=d$ contains the shot noise contribution and the second term contains the monopole contribution and the correlation function of the density fluctuations. Returning to continuous integration by decreasing the volume of cells $\Delta_c$, we have\begin{align}
        \frac{< n_{\ell m}(\mathbf{k}) n_{\ell' m'}(\mathbf{k'})>}{ \left. \bar{n} \right.^2} &= \frac{2 k k^\prime}{\pi} \int \frac{\phi(r)}{\bar{n}} j_\ell(k r) j_\ell(k' r) r^2 dr \delta_{\ell \ell'} \delta_{m m'}  \\
        &\quad + \frac{2}{\pi} \int \phi(r) k j_\ell(k r) r^2 dr\int \phi(r) k^\prime j_{\ell'}(k' r) r^2 dr \nonumber \\
        &\qquad \times   \delta_{\ell 0} \delta_{m 0} \delta_{\ell' 0} \delta_{m' 0} \nonumber \\
        &\quad + \frac{2}{\pi} \iint  \xi_g(|\mathbf{r} - \mathbf{r}'|) \phi(r) \phi(r') k k^\prime \nonumber \\ 
        &\qquad \times  j_\ell(k r) j_{\ell'}(k' r)\overline{Y_\ell^m(\mathbf{\Omega})} Y_{\ell'}^{m'} (\mathbf{\Omega'}) d\mathbf{r} d\mathbf{r}'\nonumber \;.
\end{align}
Therefore, in this expression, we recognize three terms:
\begin{itemize}
        \item the shot noise contribution, only for $l=l'$ and $m=m'$:
        \begin{equation}
                \frac{2 k k^\prime}{\pi \bar{n}}  \int \phi(r, k^{\prime}) j_\ell(k r) j_\ell(k^{ \prime} r)  r^2 \mathrm{d}r \;,
        \end{equation}
        \item the monopole contribution, only for $l=0$ and $m = 0$:
        \begin{equation}
                M_{\ell m} (k) M_{\ell' m'} (k') \;,
        \end{equation}
        \item the contribution from the power spectrum, only for $l=l'$ and $m=m'$:
        \begin{equation}
                        C_{\ell}(k, k') \;.
        \end{equation}
\end{itemize}
where $C_{\ell}(k, k')$ is defined by Eq. \eqref{eq:SFBpowspec} and $M_{l m} (k)$ can be written as\begin{equation}
\sqrt{\frac{2}{\pi}} \int \phi(r) k j_\ell(k r) r^2 dr  \delta_{\ell 0} \delta_{m 0} \;.
\end{equation}

\end{appendix}

\end{document}